\documentclass[runningheads]{svmult}

\usepackage{makeidx}   
\usepackage{graphicx}  
\usepackage{subeqnar}  
\usepackage{multicol}  
\usepackage{physprbb}  
\makeindex             

\begin{document}

\title*{Abundances and ages of the deconvolved \protect\newline 
thin/thick disks of the Galaxy}
\toctitle{Abundances and ages of the deconvolved \protect\newline 
thin/thick disks of the Galaxy}
%
%
\titlerunning{The deconvolved thin/thick disks}
%
\author{Pascal Girard
\and Caroline Soubiran
}
\authorrunning{P.Girard et C.Soubiran.}
%
%
\institute{Observatoire Aquitain des Sciences de l'Univers, L3AB, 
2 rue de l'Observatoire, BP 89, 33270 Floirac, France.}

\maketitle              

\begin{abstract}
We have investigated the abundance of several chemical elements in two large stellar
samples kinematically representative of the thin and the thick disks of the Galaxy. 
Chemical, kinematical and age data have been collected from high quality sources
in the literature. Velocities (U,V,W) have been computed and used to select stars 
with the highest probability to belong to the thin disk and the thick disk respectively.
Our results show that the two disks are chemically well separated. Both exhibit 
a decline of [$\alpha$/Fe] with increasing [Fe/H]. A transition between
the thin/thick disks stars is observed at 10 Gyr
\end{abstract}
%
%
%
A sample of 823 stars with abundances of several elements (Fe,O,Mg,Ti,Si,
Na,Ni,Al) was compiled from several papers (Ref 1 to 10) after checking the 
lack significant differences between their results. The velocities (U,V,W) 
and orbital parameters were computed for 640 Hipparcos stars having 
$\sigma_{\pi}\over \pi$$<$0.25, to make a large database combining 
kinematics and detailed abundances. Ages of 442 stars were retrieved from 
Nordstr\"{o}m et al(2004).
\begin{figure}[h!]
\begin{center}
\includegraphics[width=.30\textwidth,angle=90]{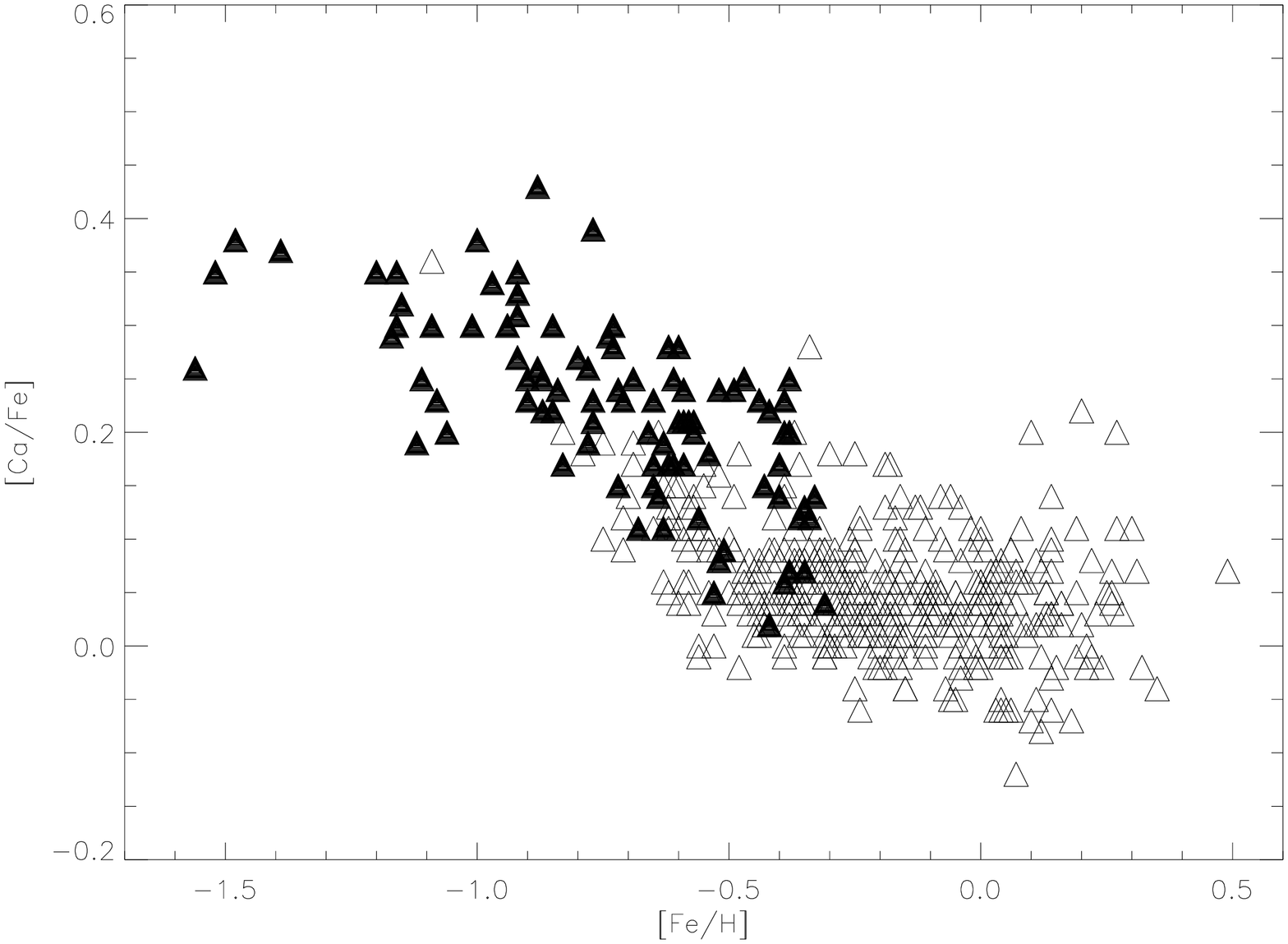}
\includegraphics[width=.30\textwidth,angle=90]{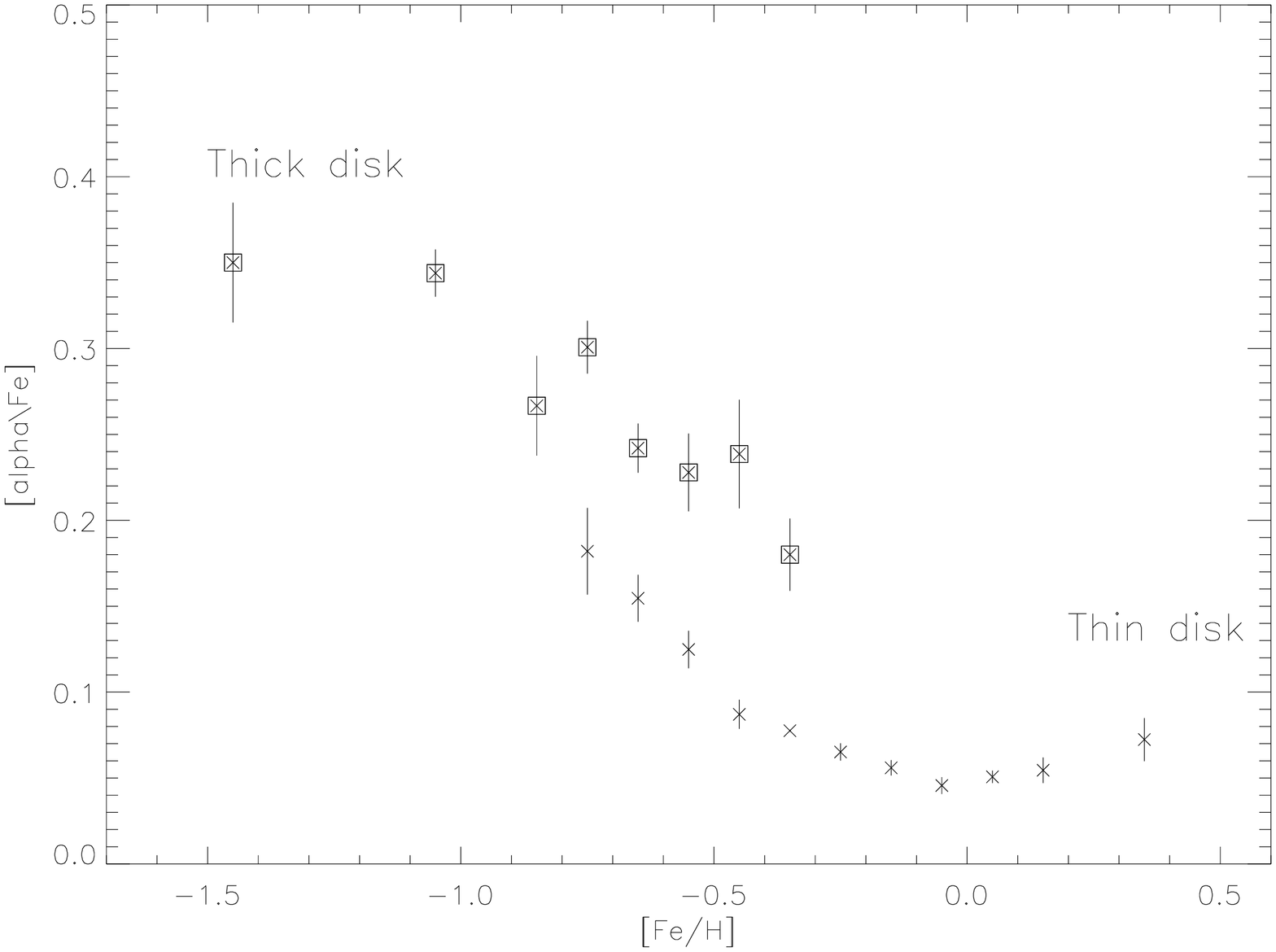}
\end{center}
\caption[]{Left : [Ca/Fe] vs [Fe/H] for the thin (empty triangles) and the thick (filled triangles) 
disk stars. Right : [$\alpha$/Fe] vs [Fe/H].}
\label{gir:F1}
\end{figure}

In order to investigate the chemical and age properties of the thin and the 
thick disks separately we have performed the deconvolution of their velocitiy
distributions. We show that about 25\% of the sample has kinematics typical of 
the thick disk, adopting for its parameters V$_{lag}$ = $-51 \mathrm{km\,s}^{-1}$
and $(\sigma_U,  \sigma_V, \sigma_W)=(63, 39, 39)  \, \mathrm{km\,s}^{-1}$. 
Stars having a probability higher than 80\% to belong to the thin and thick disks 
were selected.
Plots on Fig.1 show nicely the separation between the thin and the thick disks. 
The thick disk is $\alpha$-enhanced as compared 
to the thin disk but the decreasing trends are parallel. In the metallicity overlap, 
[$\alpha$/Fe] of the thick disk exceeds by 0.08 dex that of the thin disk. 
No clear vertical gradient of abundance in the thick disk is seen on Fig.2.
When only high precison ages (relative error $<$ 15\%) are considered, a transition between ages of 
the thin and the thick disks stars at 10 Gyr is observed (Fig.2). 
\begin{figure}[h!]
\begin{center}
\includegraphics[width=.30\textwidth,angle=90]{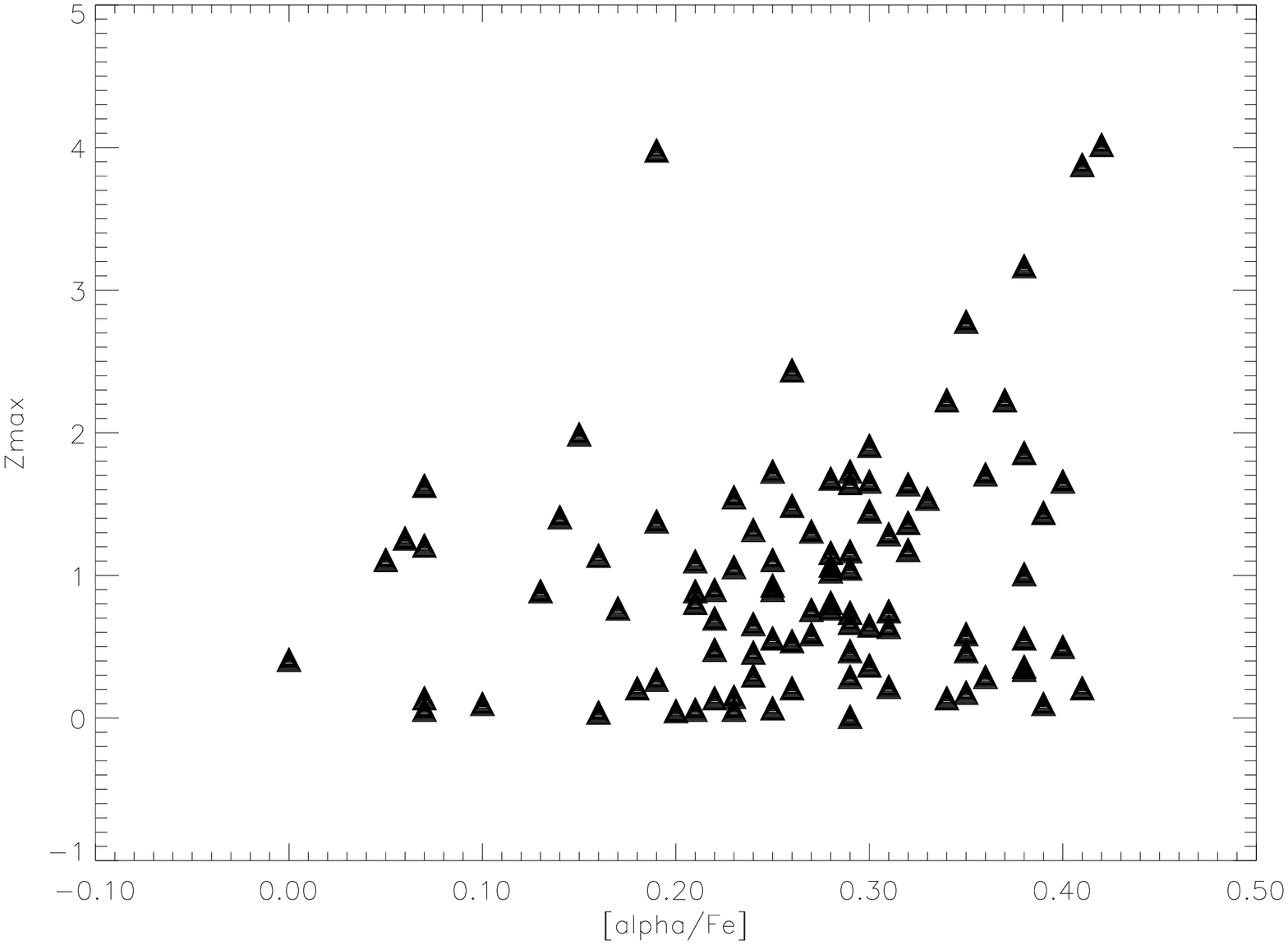}
\includegraphics[width=.30\textwidth,angle=90]{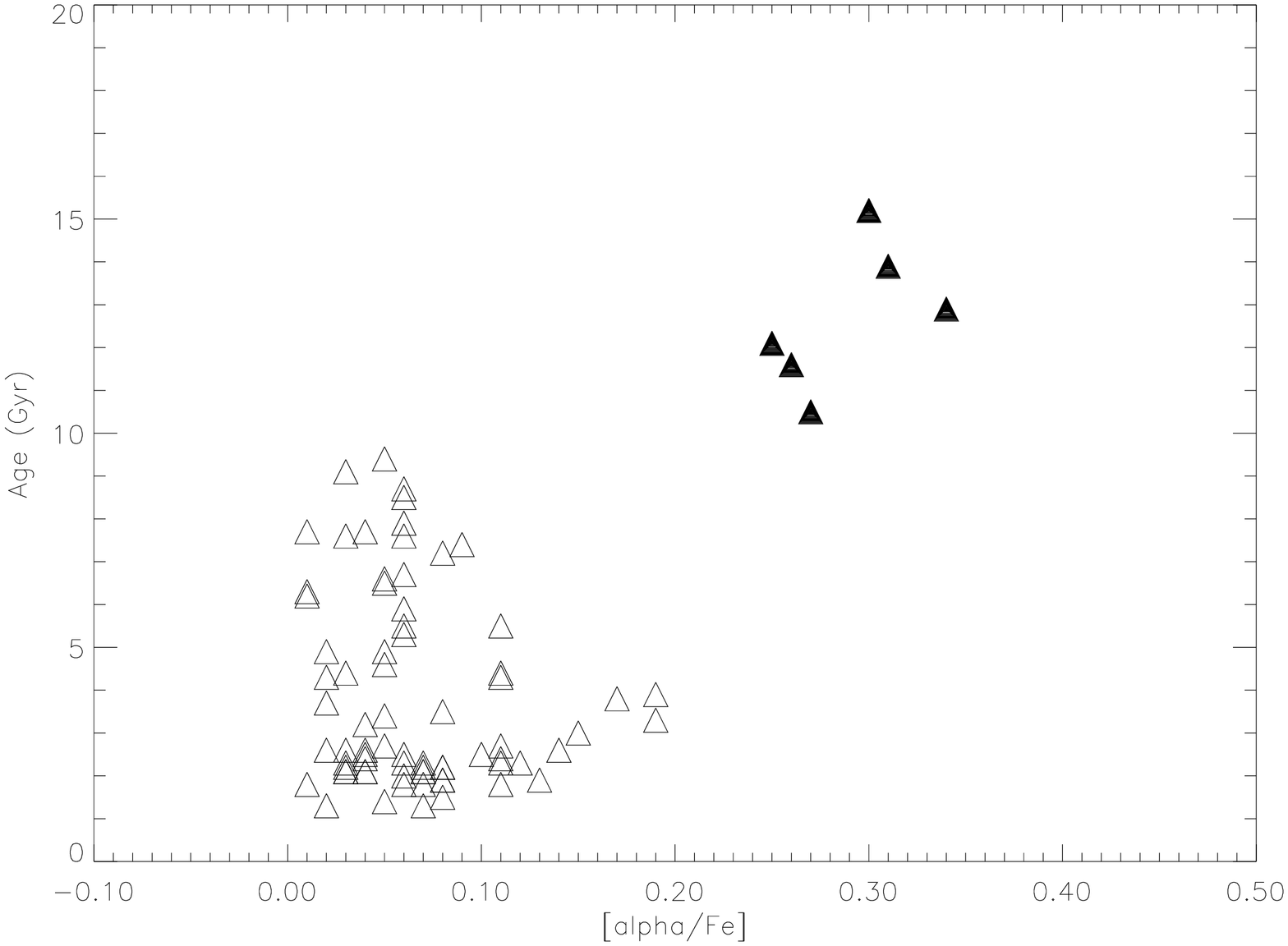}
\end{center}
\caption[]{Left : Zmax vs [$\alpha$/Fe] for the thick disk stars. 
Right : Age distribution of the thin and thick disks stars.}
\label{gir:F2}
\end{figure}
\\
{\large {\bf Conclusion.}}
Thanks to our large sample, the statistic is improved and the separation between 
the two disks is quantified. It is now clear that the thin and the thick disks are 
chemically well separated. We found a transition in the age distribution of the 
thin disk and the thick disk stars at 10 Gyr but no clear vertical gradient in the 
thick disk. These results constrain the formation scenarii of the Milky Way's 
disks.

%

\end{document}